\documentclass[epj]{svjour}
\usepackage{graphicx}
\usepackage{latexsym,amsmath,amsfonts} 
\usepackage{cite}
\begin{document}
\title{Consensus reaching in swarms ruled by a hybrid metric-topological
  distance}
\author{Yilun Shang\inst{1} \and Roland Bouffanais\inst{1}
}                     
\offprints{}          
\institute{
  Singapore University of Technology and Design,\\
  20 Dover Drive, Singapore 138682, Singapore}
\date{Received: date / Revised version: date}
%
\abstract{
Recent empirical observations of three-dimensional bird flocks and human
  crowds have challenged the long-prevailing assumption that a metric
  interaction distance rules swarming behaviors. In some cases, individual
  agents are found to be engaged in local information exchanges with a fixed
  number of neighbors, i.e. a topological interaction. However, complex system
  dynamics based on pure metric or pure topological distances both face
  physical inconsistencies in low and high density situations. Here, we
  propose a hybrid metric-topological interaction distance overcoming these
  issues and enabling a real-life implementation in artificial robotic
  swarms. We use network- and graph-theoretic approaches combined with a
  dynamical model of locally interacting self-propelled particles to study the
  consensus reaching process for a swarm ruled by this hybrid interaction
  distance. Specifically, we establish exactly the probability of reaching
  consensus in the absence of noise. In addition, simulations of swarms of
  self-propelled particles are carried out to assess the influence of the
  hybrid distance and noise.
} 
\maketitle
%

\section{Introduction}

One of the paradigmatic examples of emergent behavior in complex systems is
given by awe-inspiring collective animal behaviors such as birds flocking,
fish schooling, locusts marching, amoebae aggregating and humans
crowding~\cite{camazine,bouffanais,viscek2012,bouffanais10:_hydrod_of_cell_cell_mechan}. Over
the past two decades, research into collective phenomena has rapidly
expanded. Initially the aim was to gain insight into the elementary rules
governing such
phenomena~\cite{viscek2012,sumpter06:_princ_of_collec_animal_behav,sumpter10:_collec_animal_behav}. More
recently, following a biomimetic approach and free from some of the inherent
constraints encountered in biological systems, a host of artificial swarming
behaviors have been designed either with actual
robots~\cite{hsieh08:_decen,naruse13:_veloc_correl_swarm_robtos_direc_neigh}
or in a simulated environment~\cite{viscek2012}.

The basic mechanistic functioning of collective motion is now well understood
as being the result of multiple uncoordinated local interactions between
individuals. The central importance of these local interactions have led
scientists to experiment very many different local interaction rules, often
with the aim to reproduce fine details of some of the very specific behaviors
associated with different species of swarming
agents~\cite{viscek2012,sumpter10:_collec_animal_behav}. However, two broad
groups of local interaction rules can be discerned, each based on the
definition of a specific interaction distance thereby defining the so-called
neighborhood of interaction. The first group based on a metric distance, was
the first considered and has attracted a tremendous amount of attention (see
Ref.~\cite{viscek2012} and references therein). In the metric neighborhood
framework, each swarming agent exchanges information with all other agents
located at a fixed and given distance---assumed to be the same for all~\cite{%
hemelrijk08,hemelrijk12:_school,hemelrijk11:_some_causes_variab_shape_flock_birds}. The
metric distance was only recently challenged following the analysis of
empirical data for the dynamics of flocks of starlings~\cite{cavagna} as well
as results from the dynamics of human
crowds~\cite{ginelli10:_relev_metric_inter_flock_phenom,moussaid11:_how}. For
instance, in the topological distance framework, each and every agent
interacts with a fixed number of neighbors regardless of the distance
separating them.

Essentially, both distances are associated with distinct physiological
(resp. technological) limitations of living  (resp. artificial)
agents. Specifically, the metric neighborhood of interaction finds its origin
in the limited sensory range of individuals. Indeed, a fish in a school can
only interact with other fish it can perceive either through vision or lateral
line
sensing~\cite{coombs99:_enigm_later_line_system,bouffanais09:_hydrod_objec_recog_using_press_sensin}. On
the other hand, the topological neighborhood of interaction stems from the
limited information-processing capabilities of individuals. All living or
artificial agents possess limited cognitive and information-processing
capabilities enabling them to socially interact with a fixed number of other
agents~\cite{dusenbery92:_sensor_ecolog,emmerton93:_vision_brain_behav_birds}.
However, in real-life situations and depending on their positions within the
swarm, individuals may found themselves limited either by their sensory
apparatuses or by their internal information-processing system. Therefore, a
purely metric or purely topological distance is unable to account for this
inhomogeneity in limiting factors within the group.

Here, we address the shortcomings associated with all models based on a purely
metric or topological distance. We propose a hybrid interaction distance that
integrates both limitations in terms of sensory range as well as information
processing. Our framework is different from the hybrid metric-topological
interaction model proposed by Niizato \& Gunji~\cite{niizato11:_metric}, in
which cognitional ambiguity is accounted for by a constant switching between
class and collection cognition, thereby overcoming the problem of neighbor
selection. Very recently, a thorough comparison between the metric model and
the topological one---based on an interaction with the seven nearest neighbors---has been
carried out by Barberis \& Albano~\cite{barberis14:_eviden}. Remarkably, in
Ref.~\cite{barberis14:_eviden}, the
authors show through extensive simulations that both models share some common
features, e.g. the order parameter---scalar quantity measuring the global
consensus level within the flock---using both models has
approximately the same scaling exponent with respect to time, and the final
cluster size distributions of the flock have a similar power-law behavior.

It is worth mentioning here some influential theoretical
frameworks dealing with consensus of Vicsek-like models in the absence of
noise~\cite{tanner03:_stabl_flock_mobil_agent_part_i,tanner07:_flock_fixed_switc_networ}.
 However, these works usually
impose quite strong (sometimes unrealistic from the natural swarming standpoint) conditions on the
system. Two typical assumptions are: i) some sort of connectivity
(e.g., recurrent connectedness or the possession of spanning trees) of
the underlying interaction network, and, ii) the required balance condition, i.e.,
the out-degree equals the in-degree for each agent, when the network
is directed. As will be shown in the sequel, these restrictions do not apply
to our approach.

In the present framework,  using network- and graph-theoretic
approaches combined with the linear dynamical model by Komareji \&
Bouffanais~\cite{komareji13:_resil_contr_dynam_collec_behav}, we prove
mathematically that the achievement of global consensus is primarily
influenced by the metric component of our hybrid interaction neighborhood in
the absence of noise. Furthermore, small swarms ruled by this hybrid
interaction distance are simulated using a self-propelled particles model to
investigate the influence of three key parameters: noise level, metric radius
and number of topological neighbors within that metric radius.

\section{Hybrid metric-topological interaction distance}\label{sec:hybrid}

Consider a group of $N$ interacting agents moving in a
$\sqrt{N}\times\sqrt{N}$ square area. Each agent $i$ is described by its
velocity ${\bf v}_i=v_0\cos \theta_i\hat{x}+v_0\sin \theta_i\hat{y}$, where
$v_0$ is the constant speed and $\theta_i\in[0,2\pi)$ is the velocity
direction for $1\le i\le N$.  Therefore, the density of the agents is equal to
the unity for any size $N$ of the swarm. We assume that each agent can connect
to its at most $k$ nearest neighbors via directed information interaction
within a physical distance $R>0$. In other words, an agent is connected to at
most $k$ neighbors within a disk with radius $R$ centered about itself. Thus
the dynamical model for an individual agent $i$ can be described as
\begin{eqnarray}
  \dot{\theta}_i&=&\frac{1}{k_i}[(\theta_j-\theta_i)+(\theta_{j+1}-\theta_i)+\cdots+(\theta_{j+k_i-1}-\theta_i)]\nonumber\\
  &=&\frac{1}{k_i}(-k_i\theta_i+\theta_j+\theta_{j+1}+\cdots+\theta_{j+k_i-1}),\label{s0}
\end{eqnarray}
where $\theta_j,\theta_{j+1},\cdots,\theta_{j+k_i-1}$ represent the velocity
directions of agent $i$'s $k_i$-nearest neighbor within distance $R$ and
$k_i\le
k$~\cite{komareji13:_resil_contr_dynam_collec_behav,shang14:_influen}. The
above described $k_i$-nearest neighbor rule allows us to locally identify the
links between agents. The resulting network, through a bottom-up assembly of
the interagent links, is called the swarm signaling
network~\cite{miller,komareji13:_resil_contr_dynam_collec_behav}, for which
the specific value of $k$ has a direct impact on its connectivity character.

As stressed in our recent work~\cite{shang14:_influen}, the dynamics of this
directed swarm signaling network is intricately connected to the dynamics of
the agents in the physical space. Signaling network
structure/topology and information dynamics change on the same time scale and
are strongly interwoven. Throughout the complete dynamical process, the
signaling network maintains a constant number of nodes and some edges are
broken while new ones are being created following the hybrid interaction
rule---an agent is connected to at most $k$ neighbors within a disk with
radius $R$ centered about itself. The rate at which network edges are
changing is governed by the pace of the physical dynamics of the swarm. Hence,
contrary to the static topology of the network models considered by Aldana et 
al.~\cite{aldana07:_phase_trans_system_self_propel}, we consider here the more
general case of switching networks of interaction. Such switching events
intrinsically occur at nonuniform time intervals. As detailed in
Ref.~\cite{shang14:_influen}, we can assume without loss of generality that
those switching events are evenly distributed in time with the time interval
between switching events corresponding to the decorrelation time scale $\tau$ of the
matrix of correlations $C_{ij}=\langle \mathbf{s}_i \cdot \mathbf{s}_j
\rangle$ for the normalized velocity $\mathbf{s}_i=\mathbf{v}_i/v_0$. As all
agents move at constant speed $v_0$, the decorrelation time scale is therefore
strictly equivalent to the spatial decorrelation time scale, which given our
hybrid interaction rule is directly related to the values of $k$ and $R$.

By considering the underlying swarm signaling network, the dynamics~\eqref{s0}
of the agents is recast as
\begin{equation}
  \dot{\Theta}(t)=\mathrm{diag}(1/k_1,\cdots,1/k_N)(-L(t))\Theta(t),\label{s00}
\end{equation}
where $\Theta(t)=[\theta_1(t),\cdots,\theta_N(t)]^T$ and $L(t)$ is the
time-dependent outdegree Laplacian graph of the swarm signaling network. Our
following analysis is formally built on the framework of hybrid
$(k,R)$-nearest neighbor digraph model.

Switching networks are generally modeled using a dynamic graph $G_{s(t)}$
parameterized with a switching signal $s(t)$ that takes its values in an index
set
$\{1,\cdots,m\}$~\cite{olfati-saber07:_consen_cooper_networ_multi_agent_system}. That
is equivalent to choosing $\tau=1$, thereby imposing one specific choice of
the unit of time of the swarm dynamics. Following that approach, let
$G(N,k,R)$ denote the hybrid $(k,R)$-nearest neighbor digraph model by 
placing $N$ agents randomly and uniformly on a $\sqrt{N}\times\sqrt{N}$
square. If $R\ge\sqrt{N}$, the model reduces to the random nearest neighbor
digraph \cite{ref:eppstein,ref:balister1,ref:balister2}. We randomly choose a
sequence $G_1,G_2,\cdots,G_m,\cdots$ in $G(N,k,R)$, and generate a dynamical
random network $G(t)$ as
\begin{equation}
  G(t)=G_m,\quad \mathrm{for}\ \mathrm{any}\ t\in[m-1,m).\label{s1}
\end{equation}
Let ${\bf1}=(1,\cdots,1)^T\in\mathbb{R}^N$. We want to show the following
result

\smallskip
\noindent\textbf{Theorem A.}\quad \itshape Assume that $k\ge1$. For
all random sequences $G_1,G_2,\cdots,G_m,\cdots$ in $G(N,k,R)$, the switching
system
\begin{equation}
  \dot{\Theta}(t)=\mathrm{diag}(1/k_1,\cdots,1/k_N)(-L(t))\Theta(t)\label{s2}
\end{equation}
reaches a consensus with probability at least $1-e^{-R^2\pi}$, as
$N\rightarrow\infty$. Here $L(t)$ is the corresponding (outdegree) Laplacian
matrix of $G(t)$, and
$\Theta(t)=[\theta_1(t),\cdots,\theta_N(t)]^T$. \normalfont \smallskip

\noindent\textbf{Proof.} Since reaching consensus of system~\eqref{s2} is a
monotone increasing property with respect to the number of
edges of $G(t)$ \cite{ref:ren}, it suffices to show the case $k=1$.

Let $\xi_{ij}(t)$ be a random variable representing the directed
connection from agent $i$ to agent $j$ at time $t$. More specifically, $P(\xi_{ij}(t)=1)=(\pi
R^2k)/(N(N-1))$ and $P(\xi_{ij}(t)=0)=1-(\pi R^2k)/(N(N-1))$ for all
$i,j\in\{1,\cdots,N\}\ (i\not=j)$ and $t\ge0$. Let $M\ge1$ be an integer. By
the law of large numbers, we have
\begin{eqnarray}
&&P\left(\sum_{m=1}^M\xi_{ij}(m)\le\frac{M\pi R^2k}{2N(N-1)}\right)\nonumber\\
&=&P\left(\frac1M\sum_{m=1}^M\xi_{ij}(m)-\frac{\pi R^2k}{N(N-1)}\le-\frac{\pi R^2k}{2N(N-1)}\right)\nonumber\\
&\le&P\left(\left|\frac1M\sum_{m=1}^M\xi_{ij}(m)-\frac{\pi R^2k}{N(N-1)}\right|\ge\frac{\pi R^2k}{2N(N-1)}\right)\nonumber\\
&\le&\frac{4(N(N-1)-\pi R^2k)}{M\pi R^2k},\label{s3}
\end{eqnarray}
where we have used the variance of $\xi_{ij}(m)$ as $\pi
R^2k(N(N-1)-\pi R^2k)/N^2(N-1)^2$. Hence, we obtain
\begin{eqnarray}
&&P\left(\bigcap_{i\not=j}\left\{\sum_{m=1}^M\xi_{ij}(m)>\frac{M\pi R^2k}{2N(N-1)}\right\}\right)\nonumber\\
&=&1-P\left(\bigcup_{i\not=j}\left\{\sum_{m=1}^M\xi_{ij}(m)\le\frac{M\pi R^2k}{2N(N-1)}\right\}\right)\nonumber\\
&\ge&1-\frac{4N(N-1)(N(N-1)-\pi R^2k)}{M\pi R^2k},\label{s4}
\end{eqnarray}
which tends to 1 as $M\rightarrow\infty$.

Now define a graph $\tilde{G}$ of order $N$ whose adjacency matrix $(a_{ij})$
is given by
\begin{equation}
  a_{ij}=\left\{\begin{array}{ll}
      1,&\int_0^{\infty}\xi_{ij}(t)dt=\infty;\\
      0,&\int_0^{\infty}\xi_{ij}(t)dt<\infty.
    \end{array}\right.\label{s5}
\end{equation}
Note that $\int_0^M\xi_{ij}(t)dt=\sum_{m=1}^M\xi_{ij}(m)$. Thus,
Eq.~\eqref{s4} implies that $\tilde{G}$ is a completely connected digraph for
almost all sequences $G_1,G_2,\cdots,G_m,\cdots$.

For each agent $i$, we estimate the probability $q_k$ that there are at least
$k$ other agents within distance $R$ from it. We have
\begin{eqnarray}
  q_k&=&\sum_{j=k}^{N-1}{N\choose j}\left(\frac{\pi
      R^2}{N}\right)^j\left(1-\frac{\pi R^2}{N}\right)^{N-j}\nonumber\\
  &=&\sum_{j=k}^{N-1}\frac{\prod_{l=0}^{j-1}\left(1-\frac
      lN\right)}{j!}(\pi R^2)^je^{-\frac{\pi R^2}{N}(N-j)}\nonumber\\
  &\approx&\sum_{j=k}^{N-1}\frac{(\pi R^2)^j}{j!}e^{-\pi
    R^2},\label{s5a}
\end{eqnarray}
for large $N$. Taking $N\rightarrow\infty$, we obtain $q_1=1-e^{-\pi R^2}$. In
the following, we assume that there are at least one other agent within
distance $R$ from each agent. From the comment below Eq.~\eqref{s5}, we see
that each agent $i$ is within distance $R$ from any other agent infinitely
many times. We fix any such sequence $G_1,G_2,\cdots,G_m,\cdots$, we will show
that the consensus can be reached for the system~\eqref{s2} where
$k_1=\cdots=k_N=k=1$.

Let $\Phi(t)=[\phi_1(t),\cdots,\phi_N(t)]^T$ be a rearrangement of the vector
$\Theta(t)=[\theta_1(t),\cdots,\theta_N(t)]^T$ such that
\begin{equation}
  \phi_1(t)\le\phi_2(t)\le\cdots\le\phi_N(t).\label{s6}
\end{equation}
Note that this new vector still satisfies the equation
\begin{equation}
  \dot{\Phi}(t)=\mathrm{diag}(1/k_1,\cdots,1/k_N)(-L(t))\Phi(t)=-L(t)\Phi(t),\label{s7}
\end{equation}
except that the matrix $L(t)$ now is a result of conjugation transform made by
some permutation matrix at time $t$. In Eq.~\eqref{s7} we still write it as
$L(t)$ for simplicity. It is clear that
$\dot{\phi}_1(t)=\sum_j\xi_{1j}(\phi_j(t)-\phi_1(t))\ge0$ and
$\dot{\phi}_N(t)=\sum_j\xi_{Nj}(\phi_j(t)-\phi_N(t))\le0$. Recall from~
Eq.\eqref{s6} that $\phi_1(t)\le\phi_N(t)$. Therefore, $\phi_1(t)$ and
$\phi_N(t)$ are monotonic and bounded functions. We obtain
\begin{equation}
  \phi_1(t)\rightarrow\phi_1^*\quad \mathrm{and}\quad
  \phi_N(t)\rightarrow\phi_N^*,\label{s8}
\end{equation}
as $t\rightarrow\infty$ for some $\phi_1^*$ and $\phi_N^*$.

Define $\Psi(t)=[\psi_1(t),\cdots,\psi_N(t)]^T$. Recall that $k=1$.  Then the
outdegree of any vertex in $G(t)$ is equal to 1. Hence, it is easy to see that
there exists a diagonal matrix $B$ with diagonal elements equal to 1 or $-1$
such that
\begin{equation}
  \Psi(t)=B\Phi(t),\label{s9}
\end{equation}
and $\dot{\psi_i}(t)\ge0$ for all $i$. Since $\psi_i(t)\le|\phi_N(0)|$ (i.e.,
bounded), $\psi_i(t)$ converges for all $i$. It follows from Eq.~\eqref{s9}
that $\phi_i(t)$ also converges.  We write
\begin{equation}
  \Phi(t)\rightarrow\Phi^*=[\phi_1^*,\cdots,\phi_N^*]^T. \label{s10}
\end{equation}

Next, we claim that $\theta_i(t)$ converges for $i=1,\cdots,N$. This can be
seen as follows. Note that there exists an $\varepsilon_0>0$ such that for any
$\varepsilon<\varepsilon_0$, any pair of intervals in the family
$\{(\phi_i^*-\varepsilon,\phi_i^*+\varepsilon)\}_{i=1}^N$ is either coincident
or disjoint. For such $\varepsilon$ there exists $T>0$ such that for $t>T$,
\begin{equation}
  \{\theta_i(t)\}_{i=1}^N=\{\phi_i(t)\}_{i=1}^N\in\cup_{i=1}^N(\phi_i^*-\varepsilon,\phi_i^*+\varepsilon)\label{s11}
\end{equation}
by invoking Eq.~\eqref{s10}. Since $\theta_i(t)$ is continuous, for any
$t_1,t_2>T$ we obtain $|\theta_i(t_1)-\theta_i(t_2)|<2\varepsilon$.
Therefore, by the Cauchy convergence criterion we have
\begin{equation}
  \Theta(t)\rightarrow\Theta^*=[\theta_1^*,\cdots,\theta_N^*]^T,\label{s12}
\end{equation}
for some $\theta_i^*$ $(i=1,\cdots,N)$.

Finally, we need to show that all the above $\theta_i^*$ are equal.  From
Eq.~\eqref{s2} we have
\begin{equation}
  \dot{\theta_i}=\sum_{j=1}^N\xi_{ij}(\theta_j-\theta_i).\label{s13}
\end{equation}
In the following we will use the method of proof by contradiction.  Without
loss of generality, we assume that $\theta_{j_0}^*>\theta_{i_0}^*$. Then there
exists some $T>0$ such that
\begin{equation}
  \theta_{j_0}(t)-\theta_{i_0}(t)\ge\frac{\theta_{j_0}^*-\theta_{i_0}^*}{2}:=\delta>0\label{s15}
\end{equation}
holds for any $t>T$. Using Eqs.~\eqref{s13} and \eqref{s15} we obtain
\begin{eqnarray}
\int_T^{\infty}\xi_{i_0j_0}dt&\le&\frac{1}{\delta}\int_T^{\infty}\xi_{i_0j_0}(\theta_{j_0}-\theta_{i_0})dt\nonumber\\
&=&\frac{1}{\delta}\int_T^{\infty}\dot{\theta}_{i_0}dt\nonumber\\
&=&\frac{\theta_{i_0}^*-\theta_{i_0}(T)}{\delta}.\label{s16}
\end{eqnarray}
It then follows from the definition \eqref{s5} that $a_{i_0j_0}=0$, and hence
$\{i_0,j_0\}$ is not an edge in $\tilde{G}$. This yields a contradiction since
we know that $\tilde{G}$ is a complete digraph.  Therefore, we have
$\theta_1^*=\cdots=\theta_N^*$, which is the final consensus value of the
agents. $\Box$

As mentioned earlier, the swarm signaling network
considered in our model switches independently at each time instant
with the characteristic time scale $\tau=1$. In other words, it
follows a Markovian process of order ``zero''. More realistic models
should take into account the velocity/position co-evolution since
there are several features of self-propelled particle models that
depend crucially on the fact that motion in space is linked to local
order~\cite{chate08:_model,ginelli10:_relev_metric_inter_flock_phenom,%
gregoire03:_movin,gregoire04:_onset_of_collec_and_cohes_motion}. This simple yet tractable model is adopted here as a
very first step in understanding collective behavior ruled by a hybrid
interaction distance. A possible improvement to the model is to consider a
continuous-time Markovian system. Formally, the random network
$G(t)$ in~\eqref{s1} switches among $m$ topologies $G_1,\cdots,G_m$ in
$G(N,k,R)$, and $G(t)=G_i$ if and only if the switching signal
$s(t)=i\in\mathcal{M}:=\{1,\cdots,m\}$. The random process
$\{s(t),t\ge0\}$ is ruled by a Markov process with state space
$\mathcal{M}$ and infinitesimal generator $\Gamma=(\gamma_{ij})$
given by
\begin{multline}
\mathbb{P}(s(t+h)=j|s(t)=i)=\\
\left\{\begin{array}{cc}\gamma_{ij}h+\mathrm{o}(h), &\mathrm{when}\ s(t)\
\mathrm{jumps}\ \mathrm{from}\ i\ \mathrm{to}\
j,\\1+\gamma_{ii}h+\mathrm{o}(h),&\mathrm{otherwise}.
\end{array}\right.\nonumber
\end{multline}
Here, $\mathbb{P}$ is the concerned probability measure,
$\gamma_{ij}$ is the transition rate from state $i$ to state $j$
with $\gamma_{ij}\ge0$ if $i\not=j$,
$\gamma_{ii}=-\sum_{j\not=i}\gamma_{ij}$, and $\mathrm{o}(h)$ represents an
infinitesimal of higher order than $h$. For practical
implementation, we may set $\gamma_{ij}$ large (thus more likely) if
$G_i$ and $G_j$ differ only locally, while set $\gamma_{ij}$ small
(thus less likely) if $G_i$ and $G_j$ differ violently. The multiagent dynamical systems driven by the above Markovian
switching networks have been studied in control theory intensively
during the past few years to generate consensus behaviors; see e.g.~\cite{matei13:_conver_markov,matei08:_almos_markov,huang10:_stoch_markov}. One
of the common restrictive assumptions in these work again turns out
to be the balance condition. Clearly, these existing results do not
directly apply here.

Above we assumed that the speed $v_0$ for each agent $i$ is kept the same and
only the heading $\theta_i$ is evolving. Let $v_i^x=v_0\cos\theta_i$ and
$v_i^y=v_0\sin\theta_i$. In the following, we seek consensus of the velocity
${\bf v}_i=(v_i^x,v_i^y)$ of each agent $i$. Similarly as in Eq.~\eqref{s0},
the dynamical model for an individual agent $i$ can be described as
\begin{eqnarray}
  \dot{v}_i^\alpha&=&\frac{1}{k_i}[(v_j^\alpha-v_i^\alpha)+(v_{j+1}^\alpha-v_i^\alpha)+\cdots+(v_{j+k_i-1}^\alpha-v_i^\alpha)]\nonumber\\
  &=&\frac{1}{k_i}(-k_iv_i^\alpha+v_j^\alpha+v_{j+1}^\alpha+\cdots+v_{j+k_i-1}^\alpha),\label{s0v}
\end{eqnarray}
with $\alpha=x$ or $y$ and where
$v_j^\alpha,v_{j+1}^\alpha,\cdots,v_{j+k_i-1}^\alpha$ represent the
$\hat{\alpha}$-component of velocities of agent $i$'s $k_i$-nearest neighbors
within distance $R$ and $k_i\le k$. Denote by
$\Gamma(t)=[v_1^x(t),\cdots,v_N^x(t),v_1^y(t),\cdots,v_N^y(t)]^T$ and $L(t)$
the time-dependent outdegree Laplacian graph of the swarm signaling network as
before. The dynamics~\eqref{s0v} of the agents can be recast as
\begin{equation}
  \dot{\Gamma}(t)=(I_2\otimes\mathrm{diag}(1/k_1,\cdots,1/k_N)(-L))\Gamma(t).\label{s00v}
\end{equation}
We have the following result

\smallskip
\noindent\textbf{Theorem B.}\quad \itshape Assume that $k\ge1$. For
all random sequences $G_1,G_2,\cdots,G_m,\cdots$ in $G(N,k,R)$, the switching
system
\begin{equation}
  \dot{\Gamma}(t)=(I_2\otimes\mathrm{diag}(1/k_1,\cdots,1/k_N)(-L(t)))\Gamma(t)\label{s2v}
\end{equation}
reaches a consensus with probability at least $1-e^{-R^2\pi}$, as
$N\rightarrow\infty$. Here $L(t)$ is the corresponding (outdegree) Laplacian
matrix of $G(t)$, and $\Gamma(t)$ is given by
$\Gamma(t)=[v_1^x(t),\cdots,v_N^x(t),v_1^y(t),\cdots,v_N^y(t)]^T$.
\normalfont \smallskip

\noindent\textbf{Proof.} This result can be proved similarly as
Theorem A. Note that instead of the rearrangement~\eqref{s6} we will assume
\begin{eqnarray}
  &\phi_1^x(t)\le\phi_2^x(t)\le\cdots\le\phi_N^x(t),&\label{s6v}\\
  &\phi_1^y(t)\le\phi_2^y(t)\le\cdots\le\phi_N^y(t),&\label{s6vv}
\end{eqnarray}
where $\Phi^\alpha(t)=[\phi_1^\alpha(t),\cdots,\phi_N^\alpha(t)]^T$ with
$\alpha=x$ or $y$ are rearrangements of the two
vectors $[v_1^x(t),\cdots,v_N^x(t)]^T$ and $[v_1^y(t),\cdots,v_N^y(t)]^T$,
respectively. Now, let us set $\Phi(t)=[\Phi^x(t)^T,\Phi^y(t)^T]^T$. Similarly as in
Eq.~\eqref{s7}, we have
\begin{equation}
  \dot{\Phi}(t)=\left(\begin{array}{cc}-L_1(t)&0\\0&-L_2(t)
    \end{array}\right)\Phi(t).\label{s7v}
\end{equation}
Since the $x$-system and $y$-system are decoupled, we can prove
$v_i^x\rightarrow v^{x*}$ and $v_i^y\rightarrow v^{y*}$ ($\forall$
$i=1,\cdots,N$), respectively, for them as in the proof of Theorem A. $\Box$

\section{Self-propelled particles subjected to the hybrid interaction
  distance}

\begin{figure*}[htbp]
  \centering
  \includegraphics[width=0.8\textwidth]{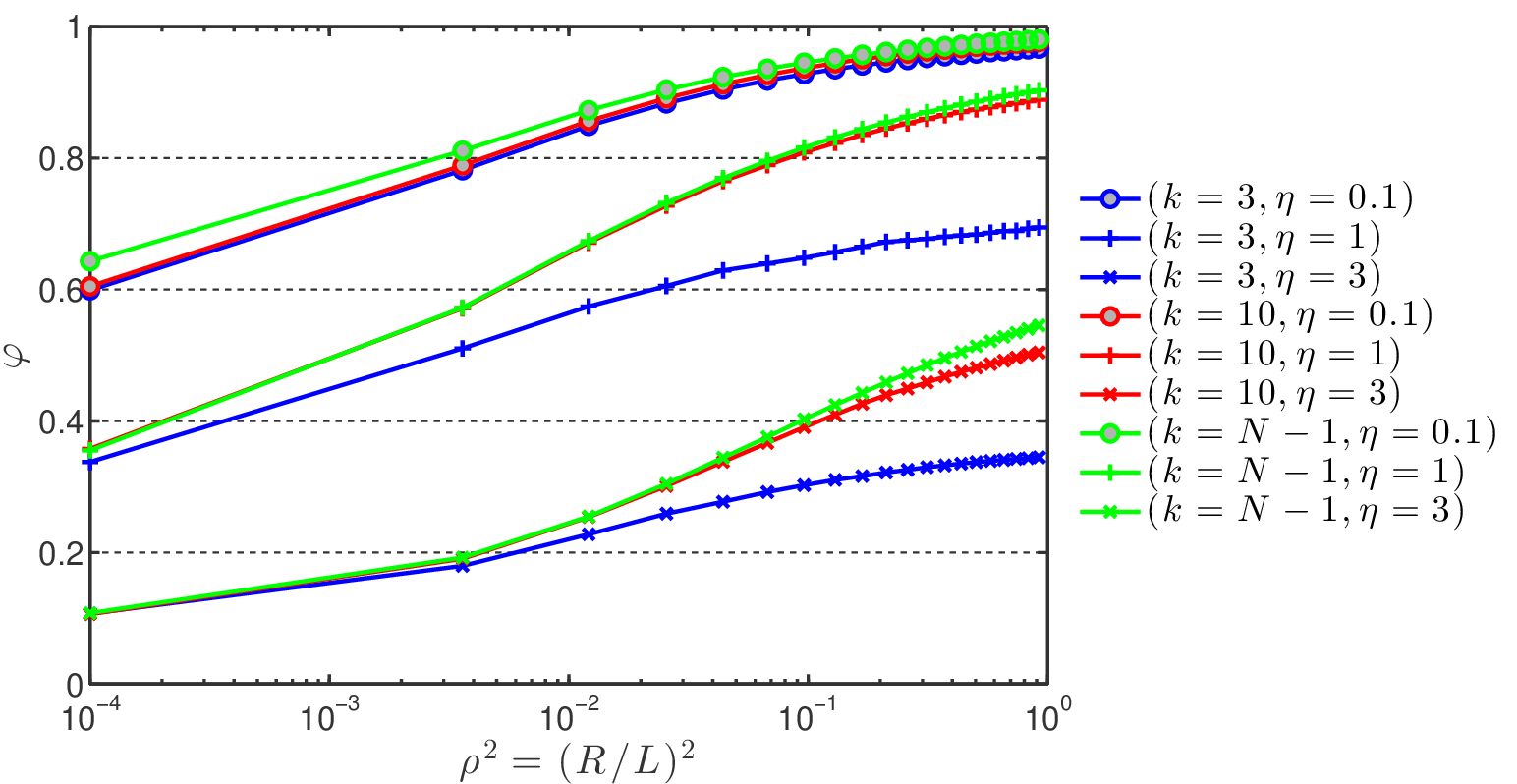}
  \caption{Order parameter $\varphi$ versus square of the normalized metric
    distance $\rho^2=(R/L)^2$ in log scale for three different values of $k$:
    $k=3,\ 10,\ N-1$, and three noise levels: low $\eta=0.1$, medium $\eta=1$,
    and high $\eta=3$. Each data point is obtained by means of a mixed
    ensemble averaging of 20 independently-computed dynamics of 500 iteration
    steps yielding a statistically ample enough
    sampling.}\label{fig:Hybrid-Radius}
\end{figure*}

Up to this point, our analysis of the swarm dynamics was limited to the ideal
case of dynamics in the absence of any noise source---stimulus and response
noises~\cite{dusenbery92:_sensor_ecolog}. With both a metric and topological
distance, it is well known that an increasing level of extraneous noise
triggers a second-order phase
transition~\cite{viscek2012,komareji13:_resil_contr_dynam_collec_behav}. For
very low noise levels, a consensus is guaranteed and after a certain transient
the swarm is globally aligned. On the contrary, beyond a certain critical
noise level no global consensus can be achieved. The consensus level within
the swarm is measured by the following order parameter
\begin{equation}\label{eq:A1}
  \varphi= \frac{1}{N} \sum\limits^{N}_{j =
    1}\frac{v_j(t)}{v_0}=\frac{1}{N} \sum\limits^{N}_{j = 1}\exp \left(\textrm{i} \theta_j(t)  \right),
\end{equation}
with $v_j$ the velocity of agent $j$ in complex notation. In the particular
case of self-propelled particles, the order parameter $\varphi$ represents the
alignment of the collective. 

At this stage it is worth highlighting that there
exists a vast range of models of self-propelled particles reported in the
literature, all stemming from the original Vicsek's
model~\cite{vicsek95:_novel}. Our objective here is not to provide an
exhaustive list of those models---many of which can be found in the following
review~\cite{viscek2012}---but instead a representative samples of some of the
key developments in relation with the present work. Alternative ways of
imposing noise have been considered in
Ref.~\cite{gregoire04:_onset_of_collec_and_cohes_motion}, and the inclusion of
cohesive forces reported in Refs.~\cite{gregoire03:_movin,chate08:_model},
while also considering the effects of the ambient fluid in
Ref.~\cite{chate08:_model}. Pure topological interactions have been reported
in Refs.~\cite{barberis14:_eviden,komareji13:_resil_contr_dynam_collec_behav,shang14:_influen},
while a topological density invariant rule based on Voronoi neighbors has been
considered in
Refs.~\cite{gregoire03:_movin,gregoire04:_onset_of_collec_and_cohes_motion},
with a particular emphasis on metric-free interactions that can be found in
Ref.~\cite{ginelli10:_relev_metric_inter_flock_phenom}. Very recently, the
effects of limited information flow due to limitations in the bandwidth of the
signaling network have been reported in Ref.~\cite{komareji14:_swarm}.

We now turn to the simulations of the dynamics of a swarm of $N=100$ agents initially evenly
distributed in a square domain of size $L=\sqrt{N}$ so as to have a unit
density as considered in Sec.~\ref{sec:hybrid}. We intentionally consider the
dynamics of small-size swarms as our interest lies in artificial
swarms~\cite{%
  hsieh08:_decen,naruse13:_veloc_correl_swarm_robtos_direc_neigh} and
some other natural collectives~\cite{calovi14:_collec,attanasi14:_infor} with
swarmers numbering in the tens to a few hundreds maximum. Given the value of
$N$ considered here, some strong small-size effects are expected and in
particular, the phase transitions---from a disordered state to collective motion
when reducing the ambient noise level---will appear to be continuous (i.e. of
second-order type) while they are actually discontinuous (i.e. of first-order
type) as proved in several studies~\cite{chate08:_collec,chate08:_model,gregoire04:_onset_of_collec_and_cohes_motion}.

For simplicity, we focus on
the constant speed case corresponding to $v_0=0.1$ that falls into the typical
range of values used for such simulations~\cite{viscek2012}. Noise can
generally be assumed to be random fluctuations with a normal
distribution~\cite{dusenbery92:_sensor_ecolog}. In the sequel, the background
noise is considered to have a normal distribution fully characterized by its
noise level, $\eta$. Specifically, starting from the continuous-time
equation~\eqref{s0}, the presence of noise leads to the following
discrete-time equation governing the dynamics of agent $i$:
\begin{align}
  \theta_i(t+\Delta t) = \theta_i(t)+\frac{\Delta t}{k_i} \left[
    (\theta_j(t)-\theta_i(t)) + \cdots \right. \nonumber \\
\left. + ( \theta_{j+k_i-1}(t)-\theta_i(t)) \right]+\Delta \theta_i,\label{smotion4}
\end{align}
where $\Delta \theta_i$ is a random number chosen with a uniform probability
from the interval $[-\eta/2,\ \eta/2]$. The agents positions $\{\mathbf{x}_i\}_{i=1,\dots,N}$ are updated
according to the discrete-time kinematic rule
\begin{equation}
\mathbf{x}_i(t+\Delta t)= \mathbf{x}_i(t) + \mathbf{v}_i \Delta t,
\end{equation}
with $\mathbf{v}_i$ having the same definition as previously introduced in
Sec.~\ref{sec:hybrid}. As for the time advance, the canonical
value $\Delta t=1$ is considered throughout~\cite{viscek2012}.The neighborhood
of interaction for this swarm model being given by the hybrid
metric-topological one, three independent parameters---namely the metric
radius $R$, the maximum number of topological neighbors $k$ within the radius
$R$ and the noise level $\eta$---influence the emergence of order.

\begin{figure*}[htbp]
  \centering
  \includegraphics[width=0.8\textwidth]{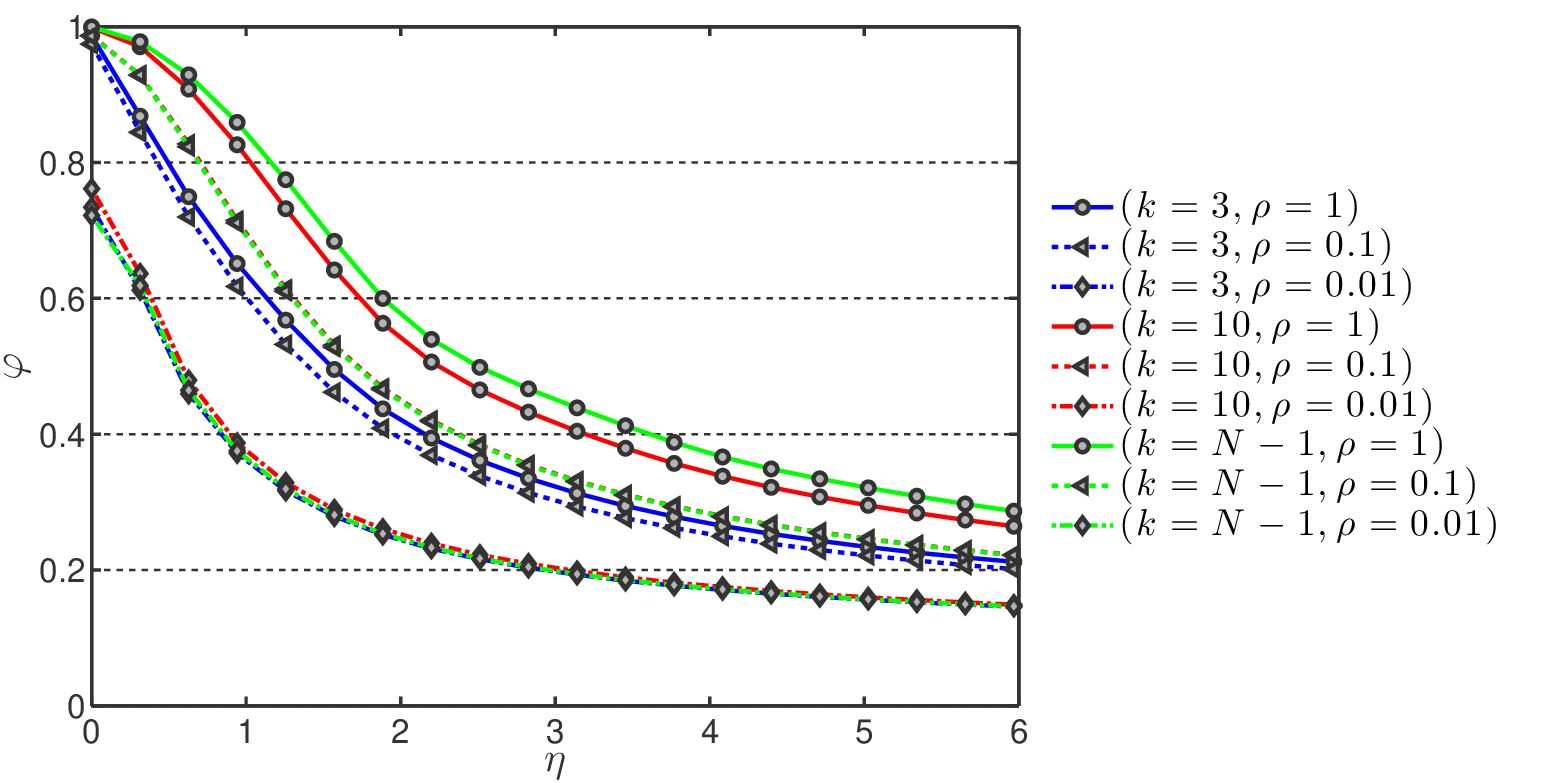}
  \caption{Order parameter $\varphi$ versus noise level $\eta$ for three
    different values of $k$: $k=3,\ 10,\ N-1$, and three vastly different
    values of the normalized metric distance: $\rho=R/L=1,\ 0.1,\ 0.01$. Each
    data point is obtained by means of a mixed ensemble averaging of 20
    independently-computed dynamics of 500 iteration steps yielding a
    statistically ample enough sampling.}\label{fig:Hybrid-Noise}
\end{figure*}

Figure~\ref{fig:Hybrid-Radius} displays the variations of the order parameter
$\varphi$ with the square of the normalized metric radius $\rho^2=(R/L)^2$ as
hinted from Theorem A. The results gathered in Fig.~\ref{fig:Hybrid-Radius}
allow us to draw several interesting comments. First, as expected from the
mathematical results in Sec.~\ref{sec:hybrid}, an increase in the metric
radius $R$ systematically yields an increase in the alignment of the swarm for
all values of $k$ and of the noise level considered. Second, for any given
value of $R$ and for all three noise levels considered, the order parameter is
found to increase with $k$ with a dramatic jump in $\varphi$ when $k$ goes
from 3 to 10 but with a minute and yet noticeable increase when $k$ goes from
10 to $N-1$. Note that the case $k=N-1$ corresponds to the extreme case of
an all-to-all connectivity of the swarm interaction network; in other words
each agent interacts with all the others which amounts to what is probably the
most cost-ineffective mode of swarming. This case $k=N-1$ also corresponds to
a pure metric interaction while the case $\rho=1$ corresponds to the purely
topological one. This second result is in complete agreement with the analysis
of the influence of the value of $k$ on the consensus reaching dynamics by
Shang \& Bouffanais~\cite{shang14:_influen}.

To further appreciate the transition between the topological and the metric
component of the hybrid interaction distance, one may compare $k$ with the
actual number $n^*(k)$ of neighbors typically found within the metric disk of radius $R$
centered about the agent. For a given value of the density $\nu=N/L^2$ of the
swarmers, one finds $n^*(k)\simeq \nu \pi R^2$. On the one hand, when
$k>n^*(k)$ it is expected that the topological component becomes ineffective
and therefore the hybrid interaction reduces to a purely metric one. Since we
investigate the constant density $\nu=1$ case, this situation is
bound to happen for small values of $R$, or equivalently $\rho$. On the
other hand, when $k<n^*(k)$, the topological component takes over the hybrid
interaction. This indeed is the case for the largest possible value for $R$, namely
$R=L$ or equivalently $\rho=1$. This analysis is consistent with the
results shown in Fig.~\ref{fig:Hybrid-Radius} and the related discussion above.

The influence of the noise level is better appreciated when turning to the
results reported in Fig.~\ref{fig:Hybrid-Noise}. An increase in the noise
$\eta$ systematically translates into a reduction of the swarm alignment. As
anticipated, the phase transition from an aligned state to a disordered one
occurs much earlier for smaller values of $R$. This observation also holds
for smaller values of $k$. However, the cases $k=10$ and $k=N-1$ yield
extremely close results for not too large values of $R$ and at any given noise level $\eta$.
Furthermore, the phase transitions associated with the rapid decay of the order
parameter $\varphi$ with noise level $\eta$, for different values of $(k,\rho)$ shown in Fig.~\ref{fig:Hybrid-Noise} are qualitatively
consistent with those reported by Barberis \&
Albano~\cite{barberis14:_eviden}. Note that the comparison can only be
qualitative owing to the fact that the time update rule used in
Ref.~\cite{barberis14:_eviden} is slightly different from the one considered
in the present study, namely
Eq.~\eqref{smotion4}. The exact same comment also applies to the comparison of
our results gathered in Fig.~\ref{fig:Hybrid-Noise} with those obtained by Niizato \& Gunji~\cite{niizato11:_metric} with yet
another time update rule and another framework for their hybrid
metric-topological interaction distance.

\section{Conclusions} \label{sec:conclusions}

A combined computational and theoretical analysis of the dynamics of swarms of
self-propelled agents subjected to a hybrid metric-topological neighborhood of
interaction has been performed. The hybrid interaction distance is devised to
overcome fundamental issues associated with inherent limitations in the
sensory range and information-processing capabilities encountered in real-life
natural and artificial complex systems. The results reported in this article
allow us to formulate the following important conclusions:
\begin{itemize}
\item[(i)] When the agent density in the field is constant (here the unity),
  the consensus dynamics is exclusively dominated by the metric component $R$
  of our hybrid model. As long as $k\geq 1$, one can always observe the
  emergence of consensus with some positive probability in the absence of
  noise. This result is further extended to the case of agents having
  non-constant speeds.
\item[(ii)] For large swarms and still in the absence of noise, the
  probability of emergence of consensus is approximately proportional to
  $R^2$.
\item[(iii)] Similarly to swarms subjected to a purely metric or a purely
  topological distance, a collapse of the swarm alignment is observed in the
  presence of a sufficiently-high noise level using our hybrid
  model. Furthermore, our results confirm the ineffectiveness of the purely
  metric model which yields a marginally higher alignment at the expense of a
  very large number of social interactions as compared to the case
  $k=10$~\cite{shang14:_influen}.
\end{itemize}
Furthermore, our initial observations reveal that the number of topological
neighbors $k$ solely affects the speed with which consensus is reached as
observed in the purely topological distance
case~\cite{shang14:_influen}. However, this last point requires a more
thorough investigation to provide a clear picture of the consensus reaching
dynamics for a swarm ruled by a hybrid interaction distance.

\section{Acknowledgments} \label{sect:acks} We thank Dr. Mohammad Komareji for
fruitful and stimulating conversations. 

This work was supported by a grant
from the SUTD-MIT International Design Centre (RB) and a grant by the
Temasek Lab (TL$@$SUTD) under the STARS project (YS \& RB).


%
\label{sect:bib}




\begin{thebibliography}{10}

\bibitem{camazine}
S.~Camazine, J.-L. Deneubourg, N.~R. Franks, J.~Sneyd, G.~Theraulaz, and
  E.~Bonabeau.
\newblock {\em Self-Organization in Biological Systems}.
\newblock Princeton University Press, Princeton, New Jersey, 2001.

\bibitem{bouffanais}
R.~Bouffanais.
\newblock {\em Design and Control of Swarm Dynamics}.
\newblock Springer, Complexity Series, Singapore, 2016.

\bibitem{viscek2012}
T.~Vicsek and A.~Zafeiris.
\newblock Collective motion.
\newblock {\em Phys. Rep.}, 517:71--140, 2012.

\bibitem{bouffanais10:_hydrod_of_cell_cell_mechan}
R.~Bouffanais and D.~K.~P. Yue.
\newblock Hydrodynamics of cell-cell mechanical signaling in the initial stages
  of aggregation.
\newblock {\em Phys. Rev. E}, 81:041920, 2010.

\bibitem{sumpter06:_princ_of_collec_animal_behav}
D.~J.~T. Sumpter.
\newblock The principles of collective animal behaviour.
\newblock {\em Phil. Trans. R. Soc. B}, 361:5--22, 2006.

\bibitem{sumpter10:_collec_animal_behav}
D.~J.~T. Sumpter.
\newblock {\em Collective Animal Behavior}.
\newblock Princeton University Press, Princeton, NJ, 2010.

\bibitem{hsieh08:_decen}
M.~Ani Hsieh, Vijay Kumar, and Luiz Chaimowicz.
\newblock Decentralized controllers for shape generation with robotic swarms.
\newblock {\em Robotica}, 26:691--701, 8 2008.

\bibitem{naruse13:_veloc_correl_swarm_robtos_direc_neigh}
Keitaro Naruse.
\newblock Velocity correlation in swarm robots with directional neighborhood.
\newblock In S.~Lee, H.S. Cho, K.J. Yoon, and J.M.~(Eds.) Lee, editors, {\em
  Intelligent Autonomous Systems 12}, pages 843--851. Advances in Intelligent
  Systems and Computing, 2013.

\bibitem{hemelrijk08}
C.~K. Hemelrijk and H.~Hildenbrandt.
\newblock Self-organised shape and frontal density of fish schools.
\newblock {\em Ethology}, 114:245--254, 2008.

\bibitem{hemelrijk12:_school}
C.~K. Hemelrijk and H.~Hildenbrandt.
\newblock Schools of fish and flocks of birds: Their shape and internal
  structure by self-organization.
\newblock {\em Interface Focus}, 2:726--737, 2012.

\bibitem{hemelrijk11:_some_causes_variab_shape_flock_birds}
C.~K. Hemelrijk and H.~Hildenbrandt.
\newblock Some causes of the variable shape of flocks of birds.
\newblock {\em PLoS one}, 6:e22479, 2011.

\bibitem{cavagna}
M.~Ballerini, N.~Cabibbo, R.~Candelier, A.~Cavagna, E.~Cisbani, I.~Giardina,
  V.~Lecomte, A.~Orlandi, G.~Parisi, A.~Procaccini, M.~Viale, and
  V.~Zdravkovic.
\newblock Interaction ruling animal collective behavior depends on topological
  rather than metric distance: Evidence from a field study.
\newblock {\em Proc. Natl. Acad. Sci. {USA}}, 105:1232--1237, 2008.

\bibitem{ginelli10:_relev_metric_inter_flock_phenom}
F.~Ginelli and H.~Chat\'e.
\newblock Relevance of metric-free interactions in flocking phenomena.
\newblock {\em Phys. Rev. Lett.}, 105:168103, 2010.

\bibitem{moussaid11:_how}
M.~Moussa\"{\i}d, D.~Helbing, and G.~Theraulaz.
\newblock How simple rules determine pedestrian behavior and crowd disasters.
\newblock {\em Proc. Natl. Acad. Sci. {USA}}, 108:6884--6888, 2011.

\bibitem{coombs99:_enigm_later_line_system}
S.~Coombs and J.~C. Montgomery.
\newblock The enigmatic lateral line system.
\newblock In R.~R. Fay and A.~N. Popper, editors, {\em Comparative Hearing:
  Fish and Amphibians, Springer Handbook of Auditory Research}, pages 319--362.
  Springer-Verlag, New York, 1999.

\bibitem{bouffanais09:_hydrod_objec_recog_using_press_sensin}
R.~Bouffanais, G.~D. Weymouth, and D.~K.~P. Yue.
\newblock Hydrodynamic object recognition using pressure sensing.
\newblock {\em Proc. R. Soc. A}, 467:19--38, 2011.

\bibitem{dusenbery92:_sensor_ecolog}
David~B. Dusenbery.
\newblock {\em Sensory Ecology: How organisms acquire and respond to
  information}.
\newblock W. H. Freeman and Co., New York, 1992.

\bibitem{emmerton93:_vision_brain_behav_birds}
J.~Emmerton and J.~Delius.
\newblock {\em Vision, Brain, and Behavior in Birds}, chapter Beyond sensation:
  Visual cognition in pigeons, pages 377--390.
\newblock MIT Press, Cambridge MA, {Z}eigler, {H}. and {B}ischof, {H}.{J}.
  edition, 1993.

\bibitem{niizato11:_metric}
T.~Niizato and Y.-P. Gunji.
\newblock Metric-topological interaction model of collective behavior.
\newblock {\em Ecol. Model.}, 222:3041--3049, 2011.

\bibitem{barberis14:_eviden}
L.~Barberis and E.~V. Albano.
\newblock Evidence of a robust universality class in the critical behavior of
  self-propelled agents: Metric versus topological interactions.
\newblock {\em Phys. Rev. E}, 89:012139, 2014.

\bibitem{tanner03:_stabl_flock_mobil_agent_part_i}
H.~G. Tanner, A.~Jadbabaie, and J.~P. Pappas.
\newblock Stable flocking of mobile agents, part i: Fixed topology.
\newblock In {\em Proc. of the 42nd IEEE Conference on Decision and Control},
  pages 2010--2015, Maui, Hawaii USA, 2003. IEEE.

\bibitem{tanner07:_flock_fixed_switc_networ}
H.~G. Tanner, A.~Jadbabaie, and J.~P. Pappas.
\newblock Flocking in fixed and switching networks.
\newblock {\em IEEE Trans. Automat. Control}, 52:863--868, 2007.

\bibitem{komareji13:_resil_contr_dynam_collec_behav}
M.~Komareji and R.~Bouffanais.
\newblock Resilience and controllability of dynamic collective behaviors.
\newblock {\em PLoS One}, 8:e82578, 2013.

\bibitem{shang14:_influen}
Y.~Shang and R.~Bouffanais.
\newblock Influence of the number of topologically interacting neighbors on
  swarm dynamics.
\newblock {\em Sci. Rep.}, 4:4184, 2014.

\bibitem{miller}
P.~Miller.
\newblock {\em The Smart Swarm}.
\newblock Penguin Group, 2010.

\bibitem{aldana07:_phase_trans_system_self_propel}
M.~Aldana, V.~Dossetti, C.~Huepe, V.~M. Kenkre, and H.~Larralde.
\newblock Phase transitions in systems of self-propelled agents and related
  network models.
\newblock {\em Phys. Rev. Lett.}, 98:095702, 2007.

\bibitem{olfati-saber07:_consen_cooper_networ_multi_agent_system}
R.~Olfati-Saber, J.~A. Fax, and R.~M. Murray.
\newblock Consensus and cooperation in networked multi-agent systems.
\newblock {\em Proc. {IEEE}}, 95(1):215--233, 2007.

\bibitem{ref:eppstein}
D.~Eppstein, M.~S. Paterson, and F.~F. Yao.
\newblock On nearest-neighbor graphs.
\newblock {\em Discrete and Computational Geometry}, 17:263--282, 1997.

\bibitem{ref:balister1}
P.~Balister, B.~Bollob\'as, A.~Sarkar, and M.~Walters.
\newblock Connectivity of random $k$-nearest neighbour graphs.
\newblock {\em Adv. Appl. Probab.}, 37:1--24, 2005.

\bibitem{ref:balister2}
P.~Balister, B.~Bollob\'as, A.~Sarkar, and M.~Walters.
\newblock A critical constant for the $k$-nearest neighbour model.
\newblock {\em Adv. Appl. Probab.}, 41:1--12, 2009.

\bibitem{ref:ren}
W.~Ren and R.W. Beard.
\newblock Consensus seeking in multiagent systems under dynamically changing
  interaction topologies.
\newblock {\em IEEE Trans. Autom. Control}, 50:655--661, 2005.

\bibitem{chate08:_model}
H.~Chat\'e, F.~Ginelli, G.~Gr\'egoire, F.~Peruani, and F.~Raynaud.
\newblock Modeling collective motion: variations on the vicsek model.
\newblock {\em Eur. Phys. J. B}, 64:451--456, 2008.

\bibitem{gregoire03:_movin}
G.~Gr\'egoire, H.~Chat\'e, and Y.~H. Tu.
\newblock Moving and staying together without a leader.
\newblock {\em Physica D-nonlinear Phenomena}, 181:157--170, 2003.

\bibitem{gregoire04:_onset_of_collec_and_cohes_motion}
G.~Gr\'egoire and H.~Chat\'e.
\newblock Onset of collective and cohesive motion.
\newblock {\em Phys. Rev. Lett.}, 92:025702, 2004.

\bibitem{matei13:_conver_markov}
I.~Matei, J.~Baras, and C.~Somarakis.
\newblock Convergence results for the linear consensus problem under markovian
  random graphs.
\newblock {\em SIAM J. Control Optim.}, 51:1574--1591, 2013.

\bibitem{matei08:_almos_markov}
I.~Matei, N.~Martins, and J.~Baras.
\newblock Almost sure convergence to consensus in {M}arkovian random graphs.
\newblock In {\em Proc. of the 47th IEEE Conference on Decision and Control},
  pages 3535–--3540, Canc\'un, Mexico, 2008.

\bibitem{huang10:_stoch_markov}
M.~Huang, S.~Dey, G.~N. Nair, and J.~H. Manton.
\newblock Stochastic consensus over noisy networks with {M}arkovian and
  arbitrary switches.
\newblock {\em Automotica}, 46:1571--1583, 2010.

\bibitem{vicsek95:_novel}
T.~Vicsek, A.~Czir{\'o}k, E.~Ben-Jacob, I.~Cohen, and O.~Shochet.
\newblock Novel type of phase-transition in a system of self-driven particles.
\newblock {\em Phys. Rev. Lett.}, 75:1226--1229, 1995.

\bibitem{komareji14:_swarm}
M.~Komareji, Y.~Shang, and R.~Bouffanais.
\newblock Swarming collapse under limited information flow between individuals.
\newblock arXiv:1409.7207, September 2014.

\bibitem{calovi14:_collec}
D.~S. Calovi, U.~Lopez, P.~Schuhmacher, H.~Chat\'e, C.~Sire, and G.~Theraulaz.
\newblock Collective response to perturbations in a data-driven fish school
  model.
\newblock arXiv:1409.6430, September 2014.

\bibitem{attanasi14:_infor}
A.~Attanasi, A.~Cavagna, L.~Del~Castello, I.~Giardina, T.~S. Grigera,
  A.~Jeli\'c, S.~Melillo, L.~Parisi, O.~Pohl, E.~Shen, and M.~Viale.
\newblock Information transfer and behavioural inertia in starling flocks.
\newblock {\em Nature Phys.}, 10:691--696, 2014.

\bibitem{chate08:_collec}
H.~Chat\'e, F.~Ginelli, G.~Gr\'egoire, and F.~Raynaud.
\newblock Collective motion of self-propelled particles interacting without
  cohesion.
\newblock {\em Phys. Rev. E}, 77:046113, 2008.

\end{thebibliography}
\end{document}